\documentclass[11pt]{article}

\usepackage{graphicx}
\usepackage{amssymb, amsmath, amsfonts, amsthm, xypic, fullpage} 
\usepackage{hyperref}

\theoremstyle{plain}

\theoremstyle{definition}

\theoremstyle{remark}

\newcommand{\bnum}{\begin{enumerate}}
\newcommand{\enum}{\end{enumerate}}

\newcommand{\bitem}{\begin{itemize}}
\newcommand{\eitem}{\end{itemize}}

\DeclareMathOperator{\argmax}{\mathrm{argmax}}

\newcommand{\s}{\mathcal{S}}
\newcommand{\G}{\mathcal{G}}

\title{{\bf Modeling Systemic Risks in Financial Markets}\footnote{A version of this paper was accepted for the 2013 TCS Global TACTiCS Conference.}}
\author{Abhijnan Rej\\[3pt] \small{Tata Consultancy Services Innovation Labs}\\\small{1 Software Units Layout, Madhapur}\\\small{Hyderabad 500081, Andhra Pradesh, India}\\[3pt] \small{Email: {\tt abhijnan.rej@tcs.com}}}
\date{}
\begin{document}

\maketitle
\pagestyle{plain}

\section{Introduction}

Systemic risk to financial markets is often defined as \emph{the risk of a major
and rapid disruption in one or more of the core functions of the financial system caused by the initial failure of one or more financial firms or a segment of the financial system} (\cite{rand-hedge}, p. 3.) This widely accepted definition sets systemic risks in financial markets as a different class of risk agents in the market face, distinct from more conventional kinds of primary and secondary risks due to a position in a certain security or more generally risks that remain in one's portfolio due to mis-hedges.
\par
The financial turmoil of 2008 and its ongoing reverberations throughout the global financial system such as the debt crisis in the Eurozone, the massive deficit in the US economy and astronomical notional value of the entire derivatives markets make systemic risk an extremely timely topic of investigation for bankers, regulators as well as governments at large. At the same time, standard tools of systemic risk management or, more to the point, catastrophic risk management, have not yet been fully successful in explaining or modeling systemic risk. (It is a widely-held hope in the financial industries that the upcoming Basel III requirements will help mitigate systemic risk though it is not clear \emph{how} these new regulations will actually mitigate such risks while preserving the hundred years old structure of capital and debt markets.)
\par
In this article we discuss some of the triggers for the onset of a full-blown catastrophic breakdown of the financial markets; we will focus on newer forms of triggers that have not been adequately covered in the 2010 Dodd-Frank Wall Street Reform and Consumer Protection Act in the US markets such as the rise of ultra-high frequency trading as well as possibilities of malicious attacks on the integrity of financial systems. We will also discuss the rise of over-the-counter (OTC) markets in securities derivatives which operate outside the standard exchanges-based framework. We will report evidence from these plethora of topics as an argument that instead of systemic risk ``going down", these new problems are keeping the entire system at the edge of another large meltdown. At the end of this paper we will present ongoing work towards developing computational tools that can simulate systemic risks in the financial markets due to various stressors; towards this end, we will propose a rapid prototype of an algorithm based on network theory that we hope to develop.
\par
The paper is organized as follows: in section \ref{sec:risks} we will present evidence from the current state of operations in the financial markets that point towards a broader picture of systemic risks as more that ``mere" contagion risks (widely studied in a series of work initiated by the Bank of England after the last global financial recession, cf. \cite{hald-may, may-arin}.) We shall also `situate' these risks in a broader global matrix of risks and point towards how financial markets systemic risks have, in fact, far bigger destructive reach spanning the entire range of human-industrial activities. In section \ref{sec:algo} we present a high-level description of an algorithm that describes it in a graph-theoretic sense (reflected in the construction of the graph ${\s_i^\ast}^D$ in section \ref{sec:algo}). By way of a conclusion in section \ref{sec:conc} we point towards challenges ahead in developing this algorithm and the attendant computational tool.

\section{Newer sources of systemic risks}\label{sec:risks}
In this paper, we will focus on the following factors that initiate, contribute to or in some cases magnify systemic risks. A glaring omission is the role of hedge funds for which I refer the reader to \cite{rand-hedge}. Another major omission is the role of the intrinsic complexity of credit-based financial derivatives; we can do no better but to point out the paper by Arora et. al. explicitly written for computer scientists \cite{arora}.
\bnum
\item{Flash Crash and the Knight Capital debacle.}
\item{Microscopic black swans as contributers to market volatility?}
\item{Risks from malicious actors with interest in disrupting financial system using cyberattacks.}
\item{Networks of OTC traded derivatives dark pools.}
\item{Geopolitical unrest, sovereign defaults and their effects on the long-term stability of international financial markets.}
\enum 

\subsection{Flash Crash and other similar problems}\label{subsec:flash}
The first major problem that was noticed about the increasing automation of the financial markets was the so-called \emph{Flash Crash of 2010}. Around 2:45 pm that day, the Dow fell unusually rapidly losing over 9\% (about 1000 points) in a couple of minutes without any noticeable external information input. After a circuit-breaker at Chicago Mercantile Exchange was switched on for a few seconds, markets climbed back up gaining by 3:30 pm most of the 600 points or that were lost as traders re-calibrated their positions and models. A following investigation of the event by the SEC, the US securities market regulator blamed a large sell-order by Proctor and Gamble that caused a large number of automated traders to exit the market simultaneously in a classic display of herding behavior. From the coupled networks viewpoint, one can imagine the Flash Crash as a result of very unstable dynamics arising from coupling a network of noise traders and fundamental traders (for example, Proctor and Gamble in this case.)
\par
This was by no means was an isolated event. On August 01 2012, the share of the retail investors gaint Knight Captial fell extremely rapidly all through the day, due to some yet-to-be-known computer glitch in the beginning of the trading day. The net loss to investors due to this one extremely rare event was to the tune of 400 million US dollars.
While the exact cause of this event remains unknown, it is not unlikely that the trigger-event at the beginning of the trading day was caused by a communication glitch between all of the trading platforms of the Knight Capital group.
\subsection{Microscopic black swans and index volatility}
Another way by which systemic risk is increasing is due to the increase in number of microscopic black swan events in the markets due to high-frequency trading. In an extremely interesting study \cite{johnson}, Johnson and collaborators analyzed high-throughput millisecond-resolution stream of prices of multiple stocks between 2006--2011. Having defined a \emph{microscopic black swan} as a market event when a stock price has to tick down at least 10 times before ticking up and the price change has to be $\geq 0.8\%$ and of durations $\leq 1500$ milliseconds, they discovered 18520 such events, making it almost one per trading day! Interestingly enough, as they observed, the overall volatility of the market indices seem to be increasing with increase in the cumulative number of microscopic black swans. This suggest an intriguing explanation of systemic risk as something that is built out of these microscopic ``fractures" perhaps leading up to another large financial meltdown.
\subsection{Financial networks and cybersecurity}
A source of increasing threat to international banking is targeted cyberattacks against major US banks \cite{finkle}. While the nature of these attacks are relatively benign for now, it is not inconceivable that cyberattacks might increase the systemic risks of financial markets. A typical scenario could be a targeted attack at one of the trading platforms for a given bank and this might trigger a major market event due to a cascading set of failures. This is a classic network-of-networks problem where the driver or trigger network might be the interconnected set of trading platforms and the main network obviously a large part of the financial system that depends on it. In such cases, one has to stress-test the stability of the main financial network by perturbing the driver networks in all possible ways.
\subsection{OTC trading networks}
A major source of systemic risk is the large market for OTC derivatives that are not centrally netted. According to Nout Wellink, President of De Nederlandsche Bank and Chairman of Basel Committee on Banking Supervision, the notional value of the OTC derivatives market is over 600 trillion US dollars and absence of an central regulator has exacerbated the risk to the rest of the financial system \cite{wellink}. The problem is further compounded by the existence of ``dark pools" where buy and sell orders are masked from the public. From a network theoretic point of view, we can imagine the ``lit markets" (i.e. those that are exchanges-based with a central netting mechanism) to be coupled to these dark pools as networks. Patterson in \cite{patterson} has provided ample evidence to make the case that this coupling is further fracturing the international markets and greatly increasing systemic risks. Absent any clear map of the OTC dark pools, agents participating in them and links between these agents, the only modeling technique that could work in examining the stability of the lit markets given the darks pools is to imagine all possible configurations for these dark pools and the way stress-points in these configurations can propagate to the lit exchanges-based markets.
\subsection{Geopolitical and geoeconomic stability and the global risk matrix}
The final and perhaps the most unmanageable source of systemic risks are destabilizing geopolitical and geoeconomic events. Examples of such events include sovereign debt defaults (such as that looming large over Greece, Portugal and possibly Spain) as well as major armed conflicts (say the possibility of an Iran--Israel conflict.) Both kinds of instability greatly increase volatility in the markets and in sharp jumps. A network theoretic way of looking at these problems could again be in terms of coupled networks; say a network of agents who are likely participants in a conflict and the way the dynamics of this network affects the global financial system. Due to the deep connectivity of the latter, a shock in a geopolitical or geoeconomic network can quickly propagate and alter the dynamics of the international financial network. For example, according to the latest Global Risks Report \cite{wef-globrisks2012}, chronic fiscal imbalances can in fact trigger a major systemic financial crisis which could be further exacerbated with a failure of critical infrastructure such as the global communication network.

\section{A networks-based model of systemic risks}\label{sec:algo}

\subsection{Assumptions on the model}\label{subsec:ass}
\subsubsection{Networks paradigm} Following the increasing interest in financial systems as networks (and the general trend towards viewing problems in biology, systems engineering and social sciences in terms of large graphs), our basic assumption is that networks and graphs provide a very valuable tool for understanding systemic risk. In fact, for us, \emph{systemic risk will be described not in terms of the standard formalism of risk measures but purely in the combinatorial language of graphs}.
\subsubsection{Systemic risk and cascading failures} It is almost tautological to state that financial networks $A$ are deeply coupled to other networks $B$ (for example computer and communication networks linking financial institutions). Following the work of Buldyrev et. al. we now understand (at least at the simulations level) how critical failures cascade through coupled networks leading to both feedback and feedfoward effects that magnify the initial shock \cite{buldyrev}. Our basic premise has been that one of the keys ways of understanding systemic risks is to look at financial networks through the coupled-networks setup. As discussed earlier in this paper, there are several instances in which a financial network $A$ is driven by some (often incompletely identified and often adversarial) ``driver network" $B$. The algorithm presented has this insight built into its core.
\subsubsection{Connectivity is the Key} Modern financial institutions are extremely well connected to each other, to the point that failure for some of them may lead to the failure of the entire system. In fact, a maximally disconnected financial system (where, say, every bank is unable to lend to another) will almost certainly cause very severe liquidity crunch. This, in turn, will feed into the real economy with extremely adverse effects. Therefore the goal of the algorithm is precisely to stress-test the connectivity of arbitrary financial networks in face of adverse scenarios.
\subsubsection{``Many Worlds Hypothesis"} Most networks that drive the financial system or has the potential to increase systemic risk often remain hidden. For example, OTC markets often operate under very little regulatory controls, even more so when such markets are fully automated (see \cite{patterson} for a nice examination of this problem.) Given this, it is often impossible for the regulator to reconstruct this hidden driver network completely. Noting this as a main problem, in the algorithm presented below we have proposed looking at stress on \emph{all possible} driver networks which may affect the larger financial network of interest. At the same time, we have made provisions for the user to decide how far he or she might be willing to entertain these machine-generated hypothetical driver networks from the initial incomplete ``expert guess" (we call such networks \emph{scenarios} in the algorithm.) It is our belief that providing for a large number of alternative scenarios may help manage tail risks.

\subsection{Algorithm}\label{subsec:algo}
\underline{{\tt Given and fixed}} 
\bnum
\item{A financial network of interest $\mathcal{G}$ with $L$ nodes viewed as a connected graph, possibly with parallel edges between its vertices.}
\item{A library ${\tt Lib}$ containing sets of update rules (that is, deletion--contraction rules) for $\mathcal{G}$ given another coupled network.}
\enum

\bnum
\item{\underline{{\tt User inputs}}
\bnum
\item{A set of $N$ nodes of interest with $N < L$ and links between them. This gives us the {\bf seed driver network} $\mathcal{S}_0$.}
\item{A set of links between nodes of $\mathcal{S}_0$ and $\mathcal{G}$. Let the set of nodes of $\mathcal{S}_0$ (resp. $\mathcal{G}$) be $V(\mathcal{S}_0)$ (resp. $V(\mathcal{G})$).
Mark the nodes for these links on $\s_0$ as $\{a_1, \ldots , a_n\} \in  V(\mathcal{S}_0)$ (calling them the {\bf link nodes}.) Mathematically, this means {\bf defining} a fixed bijective map
\begin{eqnarray*}
\partial: V(\mathcal{S}_0) & \longrightarrow & V(\mathcal{G}),\\
\partial(a_i) & {=} & b_i,
\end{eqnarray*}
where the $b_i$'s are a {\bf fixed} subset of $V(\mathcal{G})$ for all $1 \leq i \leq N$.}
\item{A rational number $\epsilon$ within a predefined range of precision ({\bf tolerance}.)}
\enum
}

\item{\underline{{\tt Bining Module} }
\bnum
\item{Program  generates  ``all possible variants" on the seed driver network $\mathcal{S}_0$ which contain the points $a_1, \ldots, a_n$, the {\bf scenarios},
\[\{\mathcal{S}_1, \ldots, \mathcal{S}_N\}.\]
Construction of scenarios: we create a complete graph $K_N$ on $N$ nodes and look at all the subgraphs of $K_N$ which contains $a_1, \ldots, a_n$ as a subset of their vertices. Each such subgraph of $K_N$ with marked points $a_1, \ldots, a_n$ is a scenario. The seed driver network $\s_0$ will obviously be in the above set.}

\item{Recall that the \emph{degree distribution} of a graph is defined in the following way \cite{newman}: let $v$ be a vertex of a (possibly directed) finite graph $\Gamma$. The \emph{degree} of $v$ is the total number of edges between $v$ and other vertices of $\Gamma$.  Define $p_k$ to be the fraction of vertices in the network with degree $k$. One can obtain a plot of $p_k$ against $k$  through a histogram of degrees of all vertices of $\Gamma$. This histogram can be viewed as a probability distribution\footnote{View $p_k$ as the probability a vertex chosen uniformly random way has degree $k$.} and is called the degree distribution of $\Gamma$ and is denoted as $\textrm{degdist}(\Gamma)$. By definition, $\textrm{degdist}(\Gamma)$ is some function of degrees $\textrm{degdist}(\Gamma) = f(k)$. The case where $f(k) \sim k^{- \alpha}$ for some rational constant $\alpha$ has been extensively studied in the context of social and economic networks.
\par
In this step, the {\tt Bining Module} computes the degree distributions of each of $\s_i$ and chooses a suitable distance between $\textrm{degdist}(\s_i)$ and $\textrm{degdist}(\s_j)$ for all $1 \leq i \leq j \leq N$. Denote this distance on a suitable space of all probability distributions $\mathcal{D}$ as $|-, - |_{\mathcal{D}}$. (An equivalent way of stating this is saying that $|-, - |_{\mathcal{D}}$) is a real-valued metric on the function space $\mathcal{D}$.)}
\item{Looks at degree distributions close to the user input $\s_0$
\[ |\textrm{degdist}(\s_i) , \textrm{degdist}(\s_0)|_{\mathcal{D}} < \epsilon \quad \forall 1 \leq i \leq N.\]
}
\item{Pick all such $\s_i$ within tolerance $\epsilon$.}
\enum
}

\item{\underline{{\tt Attack Module}}
\bnum
\item{Creates links between $\G$ and all $\s_i$ picked above using the map $\partial$ defined in the {\tt User inputs}.}
\item{\underline{{\tt User calls}} one set of rules from ${\tt Lib}$ and program attacks all nodes of all of the $\s_i$ using this set of rules and list the consequent changes in $\G$.}
\item{Program picks the set of attacks that disconnects $\G$ maximally. In other words, program looks for all attacked $\s_i$ (which we will denote as $\s_i^\ast$ for all $i$) such that they solve the optimization problem
\[ \max_{\{\s_1^\ast, \ldots, \s_n^\ast\}}[\# \textrm{ of disconnected components of }\mathcal{G}]\] or equivalently
\[ \min_{\{\s_1^\ast, \ldots, \s_n^\ast\}}[\# \textrm{ of connected components of }\mathcal{G}].\] 
}

\enum
}

\item{\underline{{\tt Program outputs}}
\bnum
\item{The set of attacks on the different scenarios and the maximally disconnected $\G$.}
\item{The ${\s_i^\ast}^D$ which when attacked causes $\G$ to be maximally disconnected is the {\bf doomsday scenario}. In other words, it is the solution to the problem
\[
{\s_i^\ast}^D =  \argmax_{\{\s_1^\ast, \ldots, \s_n^\ast\}}[\# \textrm{ of disconnected components of }\mathcal{G}]. 
\]
{\bf The doomsday scenario ${\s_i^\ast}^D$ is the graph-theoretic analog of systemic risk in our framework.}}
\enum
}
\enum

\subsection{Remarks on the algorithm}\label{subsec:rem}
\bnum
\item{In our method of creating scenarios, we have started with the complete graph with $N$ nodes $K_N$. A lot of the systemic risk literature (for example \cite{hald-may}) assume financial networks to behave as Erd\"os--R\'enyi or other classes of random graphs; this is not a problem because of the obvious fact that any random graph on $N$ vertices can be viewed as a subgraph of $K_N$.
}
\item{In the {\tt Bining Module}, we have left the exact form of the metric on $\mathcal{D}$ unspecified. A simple example is furnished by the \emph{Bhattacharya distance} which, for two continuous probability distributions $p(x)$ and $q(x)$ is defined as 
\[ |p(x) , q(x)|_{\mathcal{D}} \stackrel{\textrm{def}}{=} \int_\mathbb{R} \sqrt{p(x)q(x)} dx.\]
It turns out that this is not a metric (since triangle inequality is violated.) The modified Bhattacharya distance \[\sqrt{1- \int_\mathbb{R} \sqrt{p(x)q(x)} dx}\] is indeed a metric (known as the \emph{Hellinger distance}.) One can also choose more refined (pre)metrics like the \emph{Kullback--Leibler divergence} while looking at how ``far" the distributions $p(x)$ and $q(x)$ are from one another.
\par
Furthermore in this module, we want the program to determine the best metric on $\mathcal{D}$ \emph{based} on the degree distributions of $\s_i$ and $\s_i$. There are several important theoretical issues here that has to be resolved, cf. \cite{gibbs-su} for a mathematical discussion. A practical guideline in choosing probability metrics is to take into account the comparative tail behavior of the probability densities in question.}
\item{There are a few issues that deserve remarking on for the {\tt Attack Module}. The first pertains to the creation of a library of attack rules. The challenge here is to construct update rules which (following an attack on a given node of one of the scenarios) update $\mathcal{G}$ using the method of deletion and contraction of vertices and edges of $\mathcal{G}$. The library {\tt Lib} is to contain various sets of deletion-contraction rules coupled to the local behavior of any arbitrary scenario under attack. The actual construction of this library is likely to involve domain experts.
\par
It might be the case that a scenario might disconnect $\mathcal{G}$ completely (that is, disconnect every vertex of $\mathcal{G}$.) In this case the set of disconnected components is obviously the set of all vertices of $\mathcal{G}$ which has cardinality $L$ (by assumption.) }
\item{In the {\tt Program outputs} module the $\argmax$ condition is simply a short-hand notation for saying that we pick the worst possible scenario under attack ${\s_i^\ast}^D$ (the doomsday scenario.) While such a scenario certainly exists, at this time we can not say anything about its uniqueness.}
\enum

\section{Conclusion}\label{sec:conc}
In this article, we have provided a brief summary of systemic risks for nonspecialists, especially the ones that are currently emerging (post the 2008 financial crisis). Examples of such systemic risks have included those arising from new technologies such as high-frequency algorithmic trading, the lack of regulation of OTC derivatives trading as well as more traditional risks such as those arising from geopolitical and geoeconomic instability. We have also proposed the development of an algorithm that measures systemic risk (in a certain network theoretic sense.)
\par
Developing this algorithm further requires work in several directions:
\bnum
\item{{\bf Associating risk measures to graphs} 
\\
It should also be noted that in the algorithm presented above, we have presented a graph-theoretic analog of systemic risk to a financial market $\mathcal{G}$ in terms of identifying a doomsday scenario ${\s_i^\ast}^D$ which maximally disconnects or breaks $\mathcal{G}$. Therefore it is not really a \emph{risk measure} as conventionally understood in the sense that it is a graph and not a real number or even an integer. However, we can get actual numbers from this graph-theoretic framework by attaching suitable complexity measures to graphs constructed out of the seed network. (One example of such a measure could be $\tau({\s_i^\ast}^D$) where $\tau(\textrm{graph})$ is the number of spanning trees of an arbitrary graph.) Ongoing research attempts at associating more natural measures to the doomsday scenario.
}
\item{{\bf Simulations on a prototype}
\\
We would have to develop a prototype with a fixed explicit network of interest and an explicit seed driver network. Simulations will tell us how good our theoretical model is. At the same time, this is a computationally challenging task since the problem complexity grows with $N$ and $L$.
}
\enum

\end{document}